\documentclass{article}

\usepackage[english]{babel}

\usepackage[letterpaper,top=2cm,bottom=2cm,left=3cm,right=3cm,marginparwidth=1.75cm]{geometry}

\usepackage{amsmath}
\usepackage{graphicx}
\usepackage[colorlinks=true, allcolors=blue]{hyperref}

\begin{document}

\vspace*{0.2in}

\begin{flushleft}
{\Large
\textbf{Modelling opinion dynamics during crises as complex contagion with feedback} % Please use "sentence case" for title and headings (capitalize only the first word in a title (or heading), the first word in a subtitle (or subheading), and any proper nouns).
}
\newline
\\
Junxiang Huang\textsuperscript{1*},
Mikhail Prokopenko\textsuperscript{1,2}
% Name3 Surname\textsuperscript{2,3\textcurrency},
% Name4 Surname\textsuperscript{2},
% Name5 Surname\textsuperscript{2\ddag},
% Name6 Surname\textsuperscript{2\ddag},
% Name7 Surname\textsuperscript{1,2,3*},
% with the Lorem Ipsum Consortium\textsuperscript{\textpilcrow}
\\
\bigskip
\textbf{1} Centre for Complex Systems, The University of Sydney, Sydney, New South Wales, Australia
\\
\textbf{2} Sydney Infectious Diseases Institute, The University of Sydney, Sydney, New South Wales, Australia
\\

\bigskip

% Use the asterisk to denote corresponding authorship and provide email address in note below.
* junxiang.huang@sydney.edu.au

\end{flushleft}

\section*{Abstract}
Crises and population responses can form coupled dynamical systems, with crisis conditions shaping protective behaviours and collective responses altering the crisis trajectory. Existing models rarely capture both evolving crisis conditions and the reinforcement-dependent spread of competing behaviours. We propose a complex contagion with feedback (CCF) model that couples competing complex contagion with crisis dynamics through bidirectional feedback. Agents stochastically switch between competing states according to social reinforcement and adoption complexity, which is influenced by crisis conditions and information campaigns, while population-level behavioural change feeds back into the crisis. We formulate and analyse the model for a well-mixed population, characterising its equilibria and associated stability conditions. As a case study for model validation, we integrate CCF model with a census-calibrated agent-based model of COVID-19 transmission in Australia to represent social distancing adoption and discontinuation. We compare the CCF model with an existing opinion dynamics model and find that, despite using fewer parameters, it achieves comparable performance in reproducing recurrent infection waves. The resulting framework is parsimonious, analytically tractable, and can be adapted to different types of crises.

\clearpage
\newgeometry{top=0.85in,left=1in,right=1in,footskip=0.75in}

\section*{Introduction}

Crises and population responses can form coupled dynamical systems: crisis conditions shape risk perceptions and protective behaviours, while collective responses alter the crisis trajectory. When successful responses reduce the severity of a crisis, they may also diminish incentives to maintain those responses, creating feedbacks that generate non-linear or recurrent dynamics \cite{Funk2009,Epstein2008coupled,Montefusco2023}. Infectious disease outbreaks provide a prominent example, with recurrent epidemic waves arising from interactions among behavioural change, information dynamics, seasonality, and pathogen evolution \cite{recurrent_2021,CAMPI2022112216,Milne_Carrivick_Whyatt_2022,Chang2025opinion}. More broadly, such waves represent a general feedback process in which crises shape collective responses that, in turn, reshape the crises themselves.

Modelling opinion and behaviour dynamics during crises is challenging because crisis conditions and population responses are linked by a feedback loop. Models driven only by crisis severity often overlook social reinforcement, while models based solely on social influence may ignore the changing conditions that motivate behavioural adoption and discontinuation. Capturing these dynamics requires a bidirectional coupling between crisis and behavioural processes~\cite{Grinberger2026}. Although coupled human-crisis models have been developed in domains such as human-environment systems, policy responses to shocks, and behaviour-disease interactions \cite{Farahbakhsh2024,Fast2015,Smaldino2021,Chang2019}, they generally do not address the reinforcement-dependent spread of competing behaviours characteristic of complex contagion. A parsimonious framework is therefore needed to jointly model socially reinforced behaviour change and the evolving crisis conditions to which it responds.

A promising framework for modelling opinion-driven behaviour is complex contagion, introduced by Centola and Macy~\cite{Centola2007}. Unlike simple contagion, where a single exposure can trigger adoption, complex contagion requires reinforcement from multiple social contacts before individuals adopt a behaviour or belief~\cite{Centola2007,Centola2018book}. This threshold effect captures the social risks and credibility concerns associated with novel or contested behaviours, such as political mobilisation, technology adoption, and changes in health attitudes~\cite{Granovetter1978,Centola2010}. The framework has since been applied to domains including health behaviour, innovation diffusion, collective action, and misinformation spread~\cite{Guilbeault2018review,HbertDufresne2020}. Related threshold-based cascade models~\cite{Granovetter1978,Watts2002} and bounded-confidence models~\cite{Deffuant2000,Hegselmann2002} provide complementary mechanisms for understanding how adoption thresholds and opinion similarity shape population-level opinion dynamics.

A key extension of complex contagion considers competing contagions, in which mutually exclusive behaviours or beliefs spread through the same population~\cite{10.1098/rsif.2019.0196}. Vasconcelos \emph{et al.}~\cite{10.1098/rsif.2019.0196} integrated complex contagion, opinion dynamics, evolutionary game theory, and language competition to show how contagion complexity and network structure shape outcomes ranging from consensus to persistent polarisation. Their results highlight threshold asymmetry between competing contagions as a central determinant of collective behaviour. In crisis settings, such asymmetries can reflect differences in the perceived costs, benefits, risks, or difficulty of competing responses. As these perceptions may evolve with crisis conditions, competing complex contagion provides a natural framework for modelling socially reinforced responses that remain sensitive to external change.

However, the competing complex contagion framework~\cite{10.1098/rsif.2019.0196} is not, on its  own, sufficient to capture behaviour change during crises because it did not consider feedback between behavioural adoption and evolving crisis conditions. This aspect is important when crisis dynamics and responses have mutual influence: worsening conditions may reduce the effective complexity of adopting protective behaviours, while widespread adoption may mitigate the crisis and weaken incentives to sustain those behaviours. The framework also does not explicitly account for information campaigns that alter awareness, risk perceptions, and responsiveness to competing behavioural states~\cite{Funk2009,Fast2015}.

To address this gap, we propose a complex contagion with feedback (CCF) model, inspired by Vasconcelos \emph{et al.}~\cite{10.1098/rsif.2019.0196}. In the proposed model, agents stochastically switch between two competing states according to local social reinforcement and adoption complexity. Adoption complexity is modulated by crisis conditions and information campaigns, allowing the evolving crisis to influence behavioural resilience. In turn, population-level behavioural change feeds back into the crisis dynamics. The CCF model therefore couples complex contagion with explicit crisis dynamics, capturing both socially reinforced behaviour change and the external conditions that shape it. Our first objective is to formulate and analyse the model for a well-mixed population, deriving analytical descriptions of its equilibria for parameterised family of autonomous systems and identifying conditions leading to stable states or transitions between competing behaviours.

We then implement and validate the CCF model using social distancing adoption and discontinuation during a pandemic as competing states. Social distancing is naturally modelled as a complex contagion because its costs, uncertainties, and dependence on social norms often require reinforcement from multiple contacts before adoption occurs~\cite{Granovetter1978,centolaMacy2007complex,kittel2021peers,Latkin2021}. As disease incidence and information campaigns alter the perceived costs and benefits of adoption, bidirectional feedback emerges: rising incidence promotes protective behaviour, while widespread social distancing reduces transmission, lowers perceived risk, and encourages discontinuation. Capturing this feedback is essential for understanding recurrent epidemic waves~\cite{Bedson2021,Funk2009}. By allowing epidemic conditions to influence social reinforcement while behavioural adoption feeds back into transmission, the CCF model generates co-evolving dynamics beyond those captured by behavioural or epidemiological models alone. We validate this implementation within a previously developed census-calibrated agent-based model of COVID-19 transmission in Australia~\cite{Cliff2018,Chang2022delta,Chang2023persistence,Chang2025opinion}. Here, social distancing encompasses physical distancing, mobility reduction, mask wearing, and stay-at-home behaviours that reduce transmission~\cite{Chang2022delta}. Reflecting the largely voluntary nature of these behaviours during the Omicron period, we model social distancing as behavioural adoption, with adopters representing the population engaging in one or more such measures~\cite{Chang2023persistence}.

Recent work by Chang \emph{et al.}~\cite{Chang2025opinion} coupled an opinion dynamics model to this pandemic ABM and showed that risk aversion and social peer pressure can generate fluctuating social distancing adoption and recurrent epidemic waves. While this approach advances the integration of behavioural and epidemic dynamics, it relies on numerous calibrated parameters, including those defining the agents' memory horizon, perception fatigue, and context-specific peer influence. This complexity raises challenges for parsimony, recalibration and transferability across epidemics and crisis settings. Our second objective is therefore to compare the CCF model with this baseline within the same ABM, calibrated for the COVID-19 pandemic. By using time-varying adoption complexity of the competing contagions, we test whether the CCF model can reproduce recurrent infection waves and identify the minimal behavioural mechanisms required to generate them. The pandemic application thus serves as a validation case for assessing whether the CCF model can preserve mechanistic fidelity while reducing behavioural complexity and calibration demands.

The proposed framework makes three contributions. First, it extends competing complex contagion model by coupling socially reinforced behaviour change with explicit crisis dynamics through bidirectional feedback. Second, it provides an analytically tractable mean-field formulation for analysing equilibria and transitions. Third, its integration with a calibrated pandemic ABM demonstrates applicability in realistic settings while preserving population heterogeneity and contact-network structure. Grounded in established complex contagion theory, the model combines mechanistic fidelity with parsimony through a reduced behavioural parameterisation.

\section*{Methods}

\subsection*{Complex contagion model with feedback}
\subsubsection*{Agent characteristics}

The population is divided into two broad groups: inflexible and flexible agents. Inflexible agents adopt a fixed behaviour and do not change it during the complex contagion process, although their behaviour may influence that of their neighbours. Flexible agents, by contrast, can change their behaviour in response to the behaviour of neighbouring agents, according to the complex contagion mechanisms described in the following section. In the pandemic case study considered here, these general agent characteristics are applied to social distancing: inflexible agents are either permanently compliant or permanently non-compliant, while flexible agents choose between behaving as usual and complying with social distancing measures.

\subsubsection*{Modelling protective behaviour during crises}

To model the diffusion of protective behaviour during crises, we extend the competing complex contagion framework proposed by Vasconcelos~\emph{et al.}~\cite{10.1098/rsif.2019.0196}. The population is represented by a social network in which each node corresponds to an individual and each edge represents a social contact through which behavioural influence may occur. Each individual, represented as a vertex, holds one of two mutually exclusive competing behavioural states that are related to the crisis:
$$
s_i \in {A,B},
$$
where $A$ denotes non-adoption of social distancing and $B$ denotes adoption for the pandemic case study. Let $z_i$ denote the degree of individual $i$, and let $n_i^Y$ denote the number of neighbours of $i$ who currently hold behavioural state $Y$.

Following the complex contagion formulation \cite{10.1098/rsif.2019.0196}, an individual currently in state $X$ switches to the alternative state $Y\neq X$ with probability:
\begin{equation}
p_i^{X\rightarrow Y}
=
\left(
\frac{n_i^Y}{z_i}
\right)^{\alpha_{XY}},
\qquad X,Y\in\{A,B\},\; X\neq Y .
\label{eq:individual_update}
\end{equation}

Given this probability, Bernoulli sampling determines which agents are selected to adopt or give up social distancing behaviour. The exponent $\alpha_{XY}$ measures the complexity of adopting behaviour $Y$ among individuals who currently hold behaviour $X$. When $\alpha_{XY}=1$, adoption is linear in the fraction of neighbours holding $Y$, corresponding to a voter-model-like process or a simple contagion process. When $\alpha_{XY}>1$, adoption is nonlinear and demands reinforcement from multiple neighbours, whereas when $0<\alpha_{XY}<1$, even limited exposure to neighbours holding $Y$ yields a relatively high probability of switching~\cite{10.1098/rsif.2019.0196}.

Because the model includes inflexible individuals, complete consensus on either adoption or non-adoption behaviour is not possible, and both behaviours can coexist throughout the dynamics. However, depending on the values of $\alpha_{AB}$ and $\alpha_{BA}$, the flexible population may shift towards predominantly adoption, predominantly non-adoption, or a more balanced distribution of the two behaviours.

\subsubsection*{Equilibria of the complex contagion model in well-mixed population}
In the original complex contagion model (without feedback)~\cite{10.1098/rsif.2019.0196}, the internal fixed points of the model for a very large, well-mixed population can be described by the following autonomous nonlinear differential equation:
\begin{equation}
\dot{x} \equiv \frac{dx}{dt} = x(1-x)(x^{\alpha_{BA} - 1} - (1-x)^{\alpha_{AB} - 1})
\end{equation}
This equation has two equilibria at $x=0$ and $x=1$. An additional interior equilibrium can be identified by solving the following transcendental equation:
\begin{equation}
    1 - x = x^{\frac{\alpha_{BA} - 1}{\alpha_{AB} - 1}}
\end{equation}

\subsubsection*{Influence of crisis trend and information campaign on adoption of opinions}

To incorporate the effect of recent crisis trends on behavioural adoption, we allow the complexity parameters $\alpha_{BA}$ and $\alpha_{AB}$ to vary over time. The model uses a crisis-related indicator as its external input. In the pandemic case study considered here, this indicator is reported incidence. Let $I(t)$ denote the reported incidence on day $t$, let $w$ denote a vector of weights assigned for each day during the retrospective observation window of length $h$. At the beginning of day $t$, the weighted recent trend $\Delta I_{w,h}(t)$ is measured by the average day-to-day change over the previous $h$ completed days:

\begin{equation}
\Delta I_{w,h}(t)
=
\frac{1}{h}
\sum_{\ell=1}^{h}
w_\ell
\left[
I(t-\ell)-I(t-\ell-1)
\right].
\label{eq:incidence_trend}
\end{equation}

In the simulation, we set the weight assigned to the trend on each day to 1, i.e., $w_1 = w_2 = ... = w_h = 1$. 
When the weights are equal, Equation~\ref{eq:incidence_trend} can be simplified as follows:

\begin{equation}
    \Delta I_{w,h}(t)
    =
    \frac{1}{h}
    (I(t-1) - I(t-h-1))
\label{eq:incidence_trend_simplified}
\end{equation}

The sign of $\Delta I_{w,h}(t)$ determines how the recent crisis trajectory affects the perceived complexity of behavioural change. If $\Delta I_{w,h}(t)>0$, incidence or other indicator is increasing, indicating a worsening situation. In the pandemic case, adopting social distancing becomes easier because individuals perceive a higher risk of infection and stronger justification for adoption. Therefore, the difficulty of switching from non-adoption to adoption, referred to as the adoption complexity, $\alpha_{AB}$, decreases. At the same time, abandoning social distancing becomes more difficult because non-adoption is less socially and epidemiologically acceptable under worsening conditions. Therefore, the complexity of switching from adoption to non-adoption, $\alpha_{BA}$, increases. Conversely, if $\Delta I_{w,h}(t)<0$, incidence is decreasing, perceived risk is reduced, adopting social distancing becomes more difficult, and switching to non-adoption becomes easier.

We represent this mechanism by defining time-dependent adoption complexity parameters. Let $\alpha_{BA}(0)$ and $\alpha_{AB}(0)$ denote the base complexities in the absence of a recent trend and control the sensitivity of the behavioural response to that trend. For each population, we assign a different positive value to the normalisation coefficient $\kappa$ to account for differences in population size among the datasets. This population-level calibration scalar is chosen so that the product of $\kappa$ and the population size remains approximately constant across all datasets.

The offset to the adoption complexity by pandemic incidednce trend, $\Delta\alpha_\text{incidence}(t)$, is determined as follows:
\begin{equation}
\label{eq:offset}
\Delta\alpha_\text{incidence}(t) = \kappa  \Delta I_{w,h}(t)
\end{equation}
The offset is the product of the corresponding normalisation coefficient for the population and the recent incidence trend, $\Delta I_{w,h}(t)$.

In addition, the information campaign directly affects behaviour adoption and is considered a form of policy implementation that acts as an exogenous shock or intervention in the complex contagion dynamics. Rather than directly changing the proportion of agents who adopt social distancing behaviour, as in the baseline model~\cite{Chang2023persistence}, the campaign changes agents' perceptions of the competing behaviours. Its effect also applies to the adoption complexity offset, similar to the crisis trend influence described in Equation~\ref{eq:offset}. To distinguish the information campaign-induced change from the crisis-induced change, we denote it by $\Delta\alpha_{\text{intervention}}(t)$, where $t$ is the day of the simulation. The value of $\Delta\alpha_{\text{intervention}}(t)$ defaults to $0$ and only has a non-zero value on the date when the information campaign is actually implemented, i.e. it delivers a shock to the system.

We define the numerical strength of the information campaign as $\theta$, which is an input to the model, and represent its effect on the complex contagion dynamics by $\Delta\alpha_{\text{intervention}}(t)$. The effect of this policy intervention (information campaign) can be expressed as follows:

\begin{equation}\label{eq:intervention}
\Delta\alpha_\text{intervention}(t) = \kappa \theta
\end{equation}

The influence from pandemic incidence trend $\Delta\alpha_\text{incidence}(t)$ and information campaign $\Delta\alpha_\text{intervention}(t)$ independently contribute to the change of adoption complexity. Let $\Delta\alpha(t)$ denotes offset to the adoption complexity at time $t$, is defined as follows:

\begin{equation}
    \Delta\alpha(t) = \Delta\alpha_\text{incidence}(t) + \Delta\alpha_\text{intervention}(t)
\end{equation}

The adoption complexity parameters in both directions at a given time are determined by their values after the previous update and the offset, as defined by Equation~\ref{eq:alpha_change}. At $t=0$, we set $\alpha_{AB}(0)=\alpha_{BA}(0)=1$. To preserve the positivity of the complexity parameters, we impose a hard lower bound such that no complexity parameter $\alpha$ can decrease below a specified value $\tau$, which is set to $0.1$ for the simulations presented here. The adoption complexities can then be expressed as follows:

\begin{equation}
    \label{eq:alpha_change}
    \alpha_{AB}(t) = \text{max}(\alpha_{AB}(0)-\sum_{i=1}^{t} \Delta\alpha(i) , \tau)
    ,
    \qquad
    \alpha_{BA}(t) = \text{max}(\alpha_{BA}(0)+\sum_{i=1}^{t} \Delta\alpha(i) , \tau)
\end{equation}

The individual transition probabilities therefore become time-dependent:
\begin{equation}
p_i^{B\rightarrow A}(t)
=
\left(
\frac{n_i^A}{z_i}
\right)^{\alpha_{BA}(t)},
\qquad
p_i^{A\rightarrow B}(t)
=
\left(
\frac{n_i^B}{z_i}
\right)^{\alpha_{AB}(t)}.
\label{eq:time_dependent_transition}
\end{equation}

The adoption complexities in the two directions, $\alpha_{AB}$ and $\alpha_{BA}$, are coupled because they represent transitions between two mutually exclusive competing behaviours that respond to the same external feedback. When a crisis trend changes the perceived attractiveness or acceptability of one behaviour, it simultaneously changes that of the competing behaviour. Therefore, a change that makes switching from $A$ to $B$ easier should make switching from $B$ to $A$ more difficult, and vice versa. Coupling $\alpha_{AB}$ and $\alpha_{BA}$ in opposite directions captures this reciprocal response and ensures that the external signal affects both behavioural transitions consistently.

\subsubsection*{Modelling information campaign strength}

The information campaign may promote either adoption or non-adoption. By changing how agents perceive the two competing behaviours, it alters the relative adoption complexity and their resilience to the competing opinion. Because the information campaign directly affects $\Delta\alpha(t)$, its strength, $\theta$, is represented by a numerical value comparable to the average past trend in Equation~\ref{eq:offset}. A negative campaign strength indicates messaging that shifts agents towards less strict social distancing behaviour or makes them more resilient to the pro-adoption opinion. This may occur, for example, when government communications indicate that social distancing is no longer recommended and encourage people to get back to normal life. By contrast, a positive campaign strength indicates messaging that shifts agents towards stricter protective behaviours (e.g. social distancing) or increases their resilience to the competing non-adoption opinion, such as communications that emphasise infection risks and encourage social distancing adoption.

\subsubsection*{Validation of the CCF model through integration with an pandemic agent-based model}

\begin{figure}[htbp]
    \centering
    \includegraphics[width=0.90\linewidth]{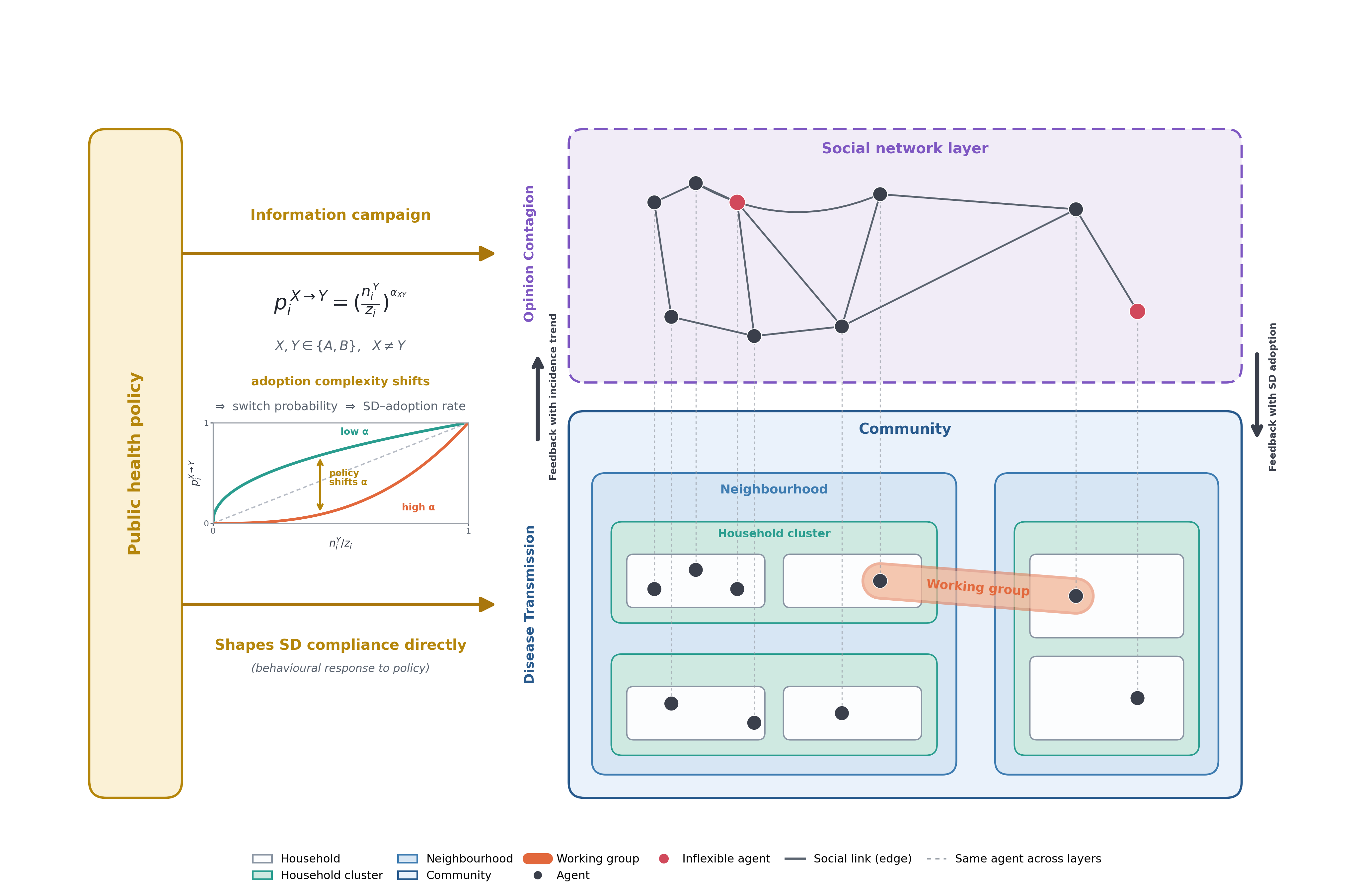}
    \caption{The overall design of the CCF model with pandemic ABM. The model have two layers: Disease transmission and Opinion Contagion, both layer shares the same set of agents, but have different forms of connections. In addition, the public health policy could have impact on both layer, through information campaign and mandates. The feedback mechanism between two layers are by the change in SD adoption rate and incidence trend}
    \label{fig:model_design}
\end{figure}

As a case study to validate the design of the Complex Contagion with Feedback (CCF) model, we pair it with a census-calibrated agent-based model (ABM) of the Australian population to simulate SARS-CoV-2 transmission alongside opinion contagion on a separately constructed social network~\cite{amtrac_2020,Nguyen2023,Chang2023persistence,Chang2025opinion,Huang2025}. The ABM is used to model infection transmission, while opinions spread in parallel on the social network. As illustrated in Figure~\ref{fig:model_design}, the model consists of two layers, the social network layer and the disease transmission layer, which share the same agents and are coupled through voluntary social-distancing adoption and epidemic incidence. Both layers can be influenced by public health policies, which can either shift the probability of adoption in the social network layer through an information campaign or directly affect SD adoption in the disease transmission layer through a mandate. The pandemic ABM serves as one application of the general CCF design, which can potentially be coupled with other crisis models. The model details and feedback mechanisms, including the population, transmission dynamics, vaccination, NPIs, and social-network construction, are described in \nameref{s1}.

\subsection*{Baseline model}

\subsubsection*{Agent Characteristics}
The baseline have similar agent characteristic configuration as our model, i.e. Inflexible Compliant, Inflexible non-Compliant, Flexible.

The main difference is that the behaviour for these two agent types balances self-evaluation and opinion dynamics. The simulation included 25\% of the population as Inflexible Compliant agents, 25\% as Inflexible Non-compliant agents, and the remaining population as Flexible agents, the proportion are exactly the same as the optimal configuration from the baseline model \cite{Chang2025opinion}.

\subsubsection*{Self-evaluation}

The perceived risk of infection of a flexible agent $i$ is defined as follows
\begin{equation}
    E^S_i(d,T) = 1-(1-\beta)^{\bar{I}(d,T)}\label{eq:SE}
\end{equation}
where $d$ is the specific day of simulation, $T$ is the time horizon (see equation \ref{eq:horizon}).

The normalised moving average of recent daily incidence is another key variable:  
\begin{equation}
    \bar{I}(d,T)=\frac{N^c}{N^n}\frac{1}{T}\sum^{T-1}_{r=0}I(d-1-r)
\end{equation}
where $\frac{N^c}{N^n}$ is the ratio between size of typical community and country population, $N^c=1000$ and $N^n=25.4 \times 10^6$, and $I(d)$ represents incidence at day $d$.

The agent memory horizon $T(d)$ represents the window size for agents to evaluate the infection risk:
\begin{equation}
    T(d)=\frac{u}{1+e^{-v(d-L)}} + T(0)\label{eq:horizon}
\end{equation}
with the parametrisation following \cite{amtrac_od_2024}: $u=28$, $v=0.25$, the day offset (delay) $L=60$ and $T(0) = 7$.

The pandemic fatigue $\gamma$ captures the individual tendency of becoming less motivated to adopt social distancing as time progresses \cite{WHO_report}. It is used to capture the fact that population adherence to stay-at-home orders and uptake of social distancing measures varies over time: an initial period of strong compliance tends to give way to considerable fatigue and seldom rebounds beyond moderate levels, thereby diminishing the effectiveness of NPIs. Notably, this fatigue is not solely the result of "contrarian" individuals, but rather manifests across broad segments of the population \cite{Chang2023persistence}.
The effect of fatigue is defined as follows:  
\begin{equation}
    \beta(d) = \beta_0(1-\gamma)^d
\end{equation}
where $\gamma$ is the perception fatigue rate, and $\beta_0=0.0001$.

\subsubsection*{Social influence}

Opinion formation at the individual level can be influenced by the views of other members within the community. Individuals may interpret and incorporate information obtained through their social contacts, leading their opinions to move closer to those held within their immediate social circles. To capture this effect, peer pressure is represented through the perceived infection risk of agent $i$, denoted by $E_i^p(d,T)$. As shown in Equation~\ref{eq:peer_pressure}, this quantity is computed as a weighted sum of the average perceived risks across the social contexts to which agent $i$ belongs.

For each social context $g \in G_i$, the model considers the average self-assessed infection risk of the other agents in that context, excluding agent $i$. The term $A_g \setminus \{i\}$ therefore denotes all members of context $g$ except agent $i$, while $E_j^s(d,T)$ represents the self-evaluated infection risk of agent $j$. The parameter $\psi_g \in [0,1]$ specifies the relative importance assigned to context $g$, with the weights over all contexts satisfying $\sum_{g \in G_i} \psi_g = 1$. Thus, Equation~\ref{eq:peer_pressure} expresses peer pressure as the weighted influence of the average opinions within the relevant social environments of the agent.

\begin{equation}
\label{eq:peer_pressure}
    E_i^p(d,T)=\sum_{g\in G_i}\psi_g\frac{1}{|A_g|-1}\sum_{j\in A_g\backslash\{i\}}E_j^s(d,T),
\end{equation}

The overall infection risk is obtained by combining the risks arising from social influence and self-evaluation:
\begin{equation}
    E_i(d)=\lambda E^P_i(d)+(1-\lambda)E^S_i(d)
\end{equation}
where the weight assigned to social influence is $\lambda=0.4$, and the corresponding weight assigned to self-evaluation is $1-\lambda$.

The flexible agent's state is determined by comparing the $E_i(d)$ to a threshold of $0.5$. If $(E_i(d) > 0.5)$, the agent is in state $B$. Otherwise, the agent is in state $A$.

\begin{equation}\label{eq:threshold}
    S(E_i(d)) =
\begin{cases}
B, & E_i(d) > 0.5 \\
A, & E_i(d) \leq 0.5
\end{cases}
\end{equation}

\subsection*{Testing dataset}

In the case study, we compare several models across multiple datasets covering the Omicron period in Australia. The models are first evaluated at the national level, after which a selection of representative states with distinct incidence patterns is examined to assess how accurately the models capture epidemiological variation with minimal parameter adjustment.

The simulations cover the period from approximately mid-November 2021 to June 2022, during which Omicron was the dominant variant in Australia. The pandemic model~\cite{Chang2025opinion} was calibrated specifically to the conditions of this period.

We begin by simulating the model using the national-level dataset of Australia and then compare the results across three states with markedly different incidence patterns: New South Wales (NSW), Victoria (VIC), and Western Australia (WA). These states are selected to evaluate the adaptability of the models.

In particular, NSW is characterised by a two-peak incidence pattern during this period. VIC exhibits a standard three-peak pattern that is characteristic of many Australian states. WA represents a unique case because it opened its borders later than the other states, resulting in a distinct pattern of recurrent epidemic waves.

\subsection*{Simulation parametrization}

For the state-level simulations, most model parameters are held constant across states, with only those variables that could reasonably differ between states adjusted during model fitting. For the baseline model, four parameters are varied: the fatigue rate $\gamma$, the perceived infection rate $\beta$, and the two parameters governing the shape of the memory horizon, namely $L$ and $u$. These parameters capture differences in pandemic severity, policy stringency, prior outbreak history, and differential desensitisation across states. For the complex contagion model, three analogous parameters are selected: the window size $h$ (days), the information campaign strength $\theta$, and the starting threshold (for the daily infection incidence) of the opinion dynamics module. These parameters functionally correspond to the memory horizon, implicit policy encoding, and perceived infection rate in the baseline model, respectively.

The asymmetry in the number of parameters, with four for the baseline model and three for the complex contagion model, reflects two considerations. First, we aim to select parameters that are as functionally comparable as possible across the two models to enable a fair comparison. Second, the complex contagion model is more parsimonious by construction, and the three selected parameters account for virtually all of its adjustable components that is applicable in the transfer setting. Full details of the parameter selection and its justification are provided in the \nameref{s3}.

In addition, for the pandemic case study, we assume that complex contagion is activated only when daily incidence exceeds a specified threshold, reflecting evidence that widespread social-distancing adoption in Australia increased only after incidence reached substantially elevated levels during the Omicron phase~\cite{Chang2023persistence}. Before this threshold is crossed, adoption is limited to a fixed proportion of inflexible agents, who subsequently seed pro-adoption opinion dynamics among the broader population. Further details are provided in the \nameref{s2}.

\section*{Results}
\subsection*{Mean-Field equilibrium analysis of the CCF model in a well-mixed population}

In this section, we analyse the complex contagion model with feedback, which is a non-autonomous version of the model introduced in previous work~\cite{10.1098/rsif.2019.0196} that incorporates coupled adoption complexities. The feedback is generated by the crisis model (a pandemic ABM in the case study) and is therefore treated as a time-dependent input. Rather than explicitly modelling the crisis model here, we examine the equilibria of the associated parameterised family autonomous systems obtained by holding the cumulative feedback offset constant at different values here.

We first consider the model without inflexible agents. Let $x\in[0,1]$ denote the fraction of the population holding opinion $A$. In the large-population limit, finite-population fluctuations become negligible, and the dynamics are described by the following non-autonomous differential equation:
\begin{equation}\label{de_no_inflexible_no_coupling}
\dot{x}
=(1-x)x^{\alpha_{BA}(t)}
-x(1-x)^{\alpha_{AB}(t)}.
\end{equation}

Under the coupled-feedback specification in Equation~\ref{eq:alpha_change}, we set $\tau = 0.1$, which gives the lower bound $\Lambda=-(1-\tau)=-0.9$ and the upper bound $\Upsilon=1-\tau=0.9$. Let $\text{s}(x, \Lambda, \Upsilon) = \max(\Lambda, \min(x, \Upsilon))$ denote the saturation function that clips $x$ to the interval $[\Lambda, \Upsilon]$.
The cumulative feedback offset applied to the system is then defined as $\rho(t) = \text{s}\left(\sum_{i=1}^{t} \Delta\alpha(i), \Lambda, \Upsilon\right)$. The adoption complexities in both directions can be expressed as follows:

\begin{equation}
\alpha_{BA}(t)=1+\rho(t),
\qquad
\alpha_{AB}(t)=1-\rho(t).
\end{equation}
Substitution into Equation~\ref{de_no_inflexible_no_coupling} gives
\begin{equation}
\dot{x}
=(1-x)x^{1+\rho(t)}
-x(1-x)^{1-\rho(t)}.
\end{equation}

To characterise the equilibrium states of the non-autonomous system under different feedback conditions, we study an auxiliary parameterised family of autonomous systems, obtained by treating $\rho$ as a constant parameter ranging over its admissible interval $[\Lambda, \Upsilon]$. We analyse the equilibria of this family as functions of $\rho$. For
$
\rho\in[\Lambda,\,\Upsilon],
$
the parameterised family of autonomous systems is:
\begin{equation}\label{eq:de_no_inflexible}
\dot{x}
=(1-x)x^{1+\rho}
-x(1-x)^{1-\rho},
\qquad
\rho\in[\Lambda,\,\Upsilon].
\end{equation}

The boundary states $x=0$ and $x=1$ are equilibria of every system in the parameterised family. For $0<x<1$, setting Equation~\ref{eq:de_no_inflexible} to zero gives
\begin{equation}\label{eq:no_inflexible_equilibrium}
\bigl[x(1-x)\bigr]^\rho=1.
\end{equation}
When $\rho\neq0$, Equation~\ref{eq:no_inflexible_equilibrium} implies $x(1-x)=1$, or equivalently
\begin{equation}
x^2-x+1=0.
\end{equation}
The discriminant of this quadratic equation is
\begin{equation}
\Delta=-3<0,
\end{equation}
so it has no real solutions. Therefore, no interior equilibrium exists when $\rho\neq0$.

For $\rho>0$, the net rate of change of $x$ is negative throughout the interior, and trajectories of the parameterised system move toward $x=0$. For $\rho<0$, the net rate of change of $x$ is positive throughout the interior, and trajectories move toward $x=1$. When $\rho=0$, the net rate of change of $x$ vanishes identically, so every $x\in[0,1]$ is a neutral equilibrium of the parameterised system.

We next consider the case with inflexible agents. Let $Z$ denote the fraction of the total population inflexibly committed to each opinion, so that the total inflexible fraction is $2Z$. Let $x\in[0,1-2Z]$ denote the fraction of the total population comprising flexible agents who currently hold opinion $A$. The total fraction holding opinion $A$, including inflexible agents, is therefore
\begin{equation}
x_{\mathrm{full}}=x+Z.
\end{equation}
The corresponding mean-field dynamics are
\begin{equation}\label{eq:analytical_alpha}
\dot{x}
=(1-x-2Z)(x+Z)^{\alpha_{BA}(t)}
-x(1-x-Z)^{\alpha_{AB}(t)}.
\end{equation}

As in the case without inflexible agents, we characterise the equilibria using a parameterised family of autonomous systems, with the cumulative feedback offset $\rho$ treated as a constant parameter varying over its admissible range. This together with setting $\dot{x}=0$ gives the following transcendental equation:
\begin{equation}\label{eq:equivalence}
(1-x^*-2Z)(x^*+Z)^{1+\rho}
=x^*(1-x^*-Z)^{1-\rho},
\end{equation}
where $x^*$ denotes the equilibrium fraction of flexible agents holding opinion $A$ in the autonomous system parameterised by $\rho$.

\begin{figure}[htbp]
    \centering
    \includegraphics[width=0.75\linewidth]{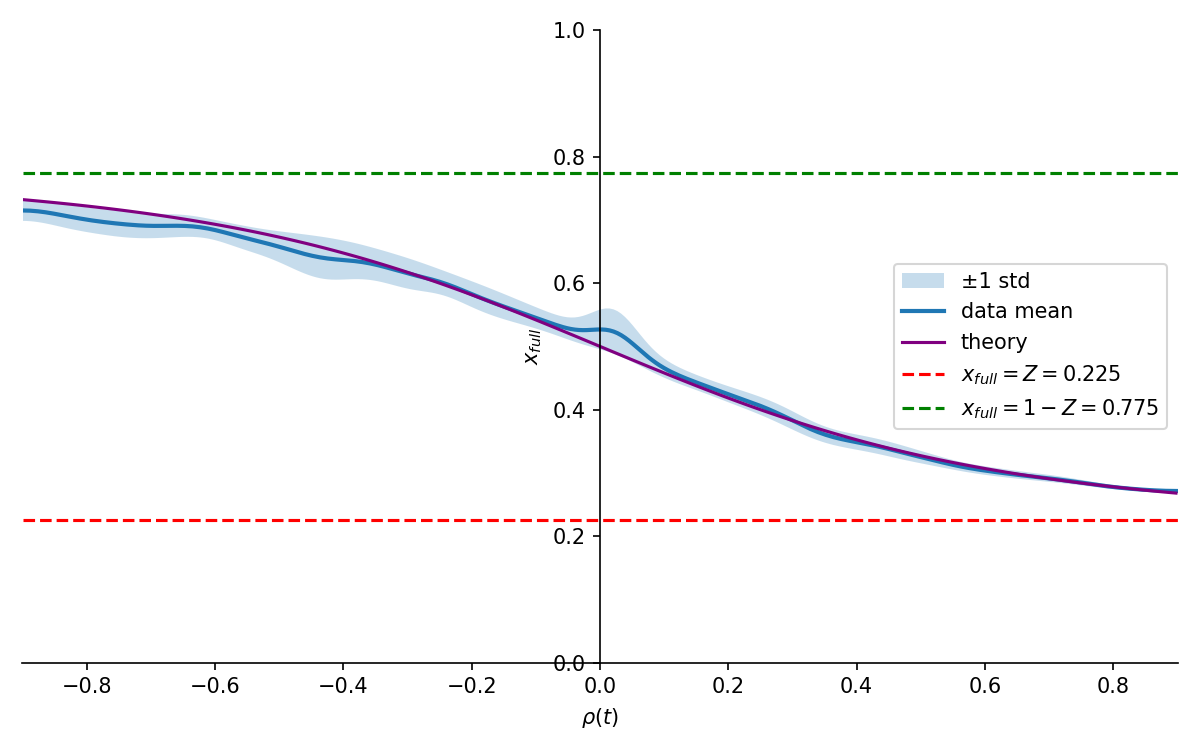}
    \caption{Comparison between the analytical equilibrium proportion of SD adopters in a well-mixed population and the CCF simulation results obtained using the WA dataset. The simulation data consist of daily cumulative adoption-complexity offsets and SD adoption proportions collected from 90 simulation runs.}
    \label{fig:adoption_theoretical_actual}
\end{figure}

Equation~\ref{eq:equivalence}, with $Z = 0.225$, is solved numerically over the admissible range of $\rho$ to obtain the equilibrium curve for the parameterised family of autonomous systems, the result is shown in Figure~\ref{fig:adoption_theoretical_actual}. In addition, as Figure~\ref{fig:adoption_theoretical_actual} shows, the simulation results are broadly consistent with this equilibrium curve over $\rho\in[\Lambda,\Upsilon]$. The remaining deviations reflect finite-population stochastic fluctuations and the transient effects arising from the time-varying feedback input.

\subsection*{Pandemic simulation}

We evaluated our CCF model against the baseline model under four population configurations: Australia nationwide (National), NSW, VIC, and WA. The best-fit parameter configuration for each population is reported in Table~\ref{tab:parameter_baseline} for the baseline model and Table~\ref{tab:parameter_contagion} for the complex contagion model.

For the baseline model, the parameter configuration for the national-level data follows the optimal configuration used in the baseline model (PP3) reported by Chang~\emph{et al.}~\cite{Chang2025opinion}. In contrast, all state-level datasets require some degree of parameter tuning relative to the national-level configuration to achieve a better fit to the corresponding state-level incidence data. As shown in Table~\ref{tab:parameter_baseline}, NSW and VIC require only minor parameter shifts to achieve optimal performance. This is expected because both states contribute substantially to national incidence, making the aggregate national-level parameter configuration relatively close to their state-level optima. By contrast, WA requires a more substantial adjustment from the national-level configuration. This may be attributed to WA's relatively smaller population and its markedly different policy settings and pandemic conditions piror and during the study period.

For the complex contagion model, Table~\ref{tab:parameter_contagion} presents three free parameters. For the starting threshold, the National population requires the largest value, whereas NSW and VIC share a smaller starting threshold. WA represents a special case, requiring a much smaller starting threshold to achieve optimal simulation accuracy. This may be related to the distinct pandemic history and circumstances in WA, where stricter restrictions were maintained and reopening occurred significantly later during the Omicron period. The optimal incidence window size also varies across populations, with values of 28 days for the National population, 42 days for both NSW and VIC, and 14 days for WA.

The information-campaign parameter for this case stuydy is defined as the strength of an additional, explicitly modelled campaign intervention. At the state level, this parameter can be assigned according to the timing and implementation of campaigns within each state. In contrast, the national dataset aggregates states with heterogeneous campaign policies, implementation times, and intervention strengths. Consequently, there is no single national-level campaign value that can be applied consistently without imposing an artificial assumption of a uniform nationwide intervention. We therefore set the parameter to zero in the national-level analysis. This zero value denotes the absence of an additional national-level campaign intervention in the model, rather than the absence of information campaigns or their effects across the country. For NSW and VIC, the optimal information campaign strength values are identical, which is consistent with the broadly synchronised information-campaign-related policy settings observed in these two states. WA, however, requires a stronger information campaign impact, reflecting the sharper transition in government recommendations from stringent NPIs to reopening.

An interesting observation from the parameter configuration of the complex contagion model (Table~\ref{tab:parameter_contagion}) is that NSW and VIC share exactly the same optimal configuration. This finding is consistent with the relatively synchronised policy changes and similar pandemic circumstances observed in these two states compared with WA.

\begin{table}[htbp]
    \centering
    \caption{Tuned optimal free parameters for baseline and complex contagion models across datasets. Memory horizon is abbreviated as MH.}
    \label{tab:parameter_baseline}
    \begin{tabular}{|l|c|c|c|c|}
        \hline
        & \multicolumn{4}{c|}{\textbf{Baseline Model}}  \\ \hline
        \cline{2-5}
        \textbf{Dataset} & \textbf{Fatigue Rate $\gamma$} & \ \textbf{MH $u$} & \textbf{MH $L$} & \textbf{Risk Perception $\beta$} \\
        \hline
        National & 0.002 & 28 & 60 & 0.5 \\ \hline
        NSW & 0.001 & 35 & 60 & 0.5 \\ \hline
        VIC & 0.001 & 28 & 60 & 0.5 \\ \hline
        WA & 0.010 & 28 & 100 & 0.9 \\
        \hline
    \end{tabular}
\end{table}

\begin{table}[]
    \centering
    \caption{Tuned optimal free parameters for baseline and complex contagion models across datasets. $h$ denotes the length of the observation window, and $\theta$ denotes the strength of the information campaign.}
    \label{tab:parameter_contagion}
    \begin{tabular}{|l|c|c|c|}
        \hline
        & \multicolumn{3}{c|}{\textbf{Complex Contagion Model}} \\
        \cline{2-4}
        \textbf{Dataset} \ & \textbf{Starting threshold} & $h$ & $\theta$ \\
        \hline
        National  & 50000 & 28 & 0 \\ \hline
        NSW  & 20000 & 42 & -1500 \\ \hline
        VIC  & 20000 & 42 & -1500 \\ \hline
        WA  & 500 & 14 & -3000 \\
        \hline
    \end{tabular}
\end{table}

\begin{table}[htbp]
    \centering
    \caption{Comparison of NRMSE between Baseline and New Model across Datasets, with the best configuration}
    \label{tab:nrmse_comparison}
    \begin{tabular}{|l|c|c|c|}
        \hline
        \textbf{Dataset} & \textbf{Baseline (\%)} & \textbf{Complex Contagion Model (\%)} & \textbf{Improvement (\%)} \\
        \hline
        National & 39.53 & 35.43 & 4.10 \\ \hline
        NSW &  44.80 & 50.51 & -5.71 \\ \hline
        VIC & 49.52 & 38.71 & 10.81 \\ \hline
        WA (125days) & 57.75 & 24.51 & 32.24 \\
        \hline
    \end{tabular}
\end{table}

\begin{table}[htbp]
    \centering
    \caption{Comparison of Directional Accuracy (higher is better) between Baseline and New Model across Datasets, with the same best configuration choosen according to the NRMSE}
    \label{tab:mda_comparison}
    \begin{tabular}{|l|c|c|c|}
        \hline
        \textbf{Dataset} & \textbf{Baseline (\%)} & \textbf{Complex Contagion Model (\%)}\\
        \hline
        National & 91.10 & 95.21 \\ \hline
        NSW &  90.85 & 90.14 \\ \hline
        VIC & 87.23 & 84.40 \\ \hline
        WA (125days) & 87.21 & 89.53 \\
        \hline
    \end{tabular}
\end{table}

\subsection*{Overview of free parameter effects}

Both models include a parameter that controls the starting point of the opinion dynamics process: the starting threshold for the CCF model and the risk perception $\beta$ for the baseline model. The starting threshold in the CCF model primarily affects the magnitude of the first wave, such that a higher starting threshold results in a lower first wave, whereas a lower starting threshold results in a higher first wave. The effect of the risk perception $\beta$ is slightly more complex. It first affects the self-evaluation value defined in Equation~\ref{eq:SE} and then, together with the threshold of adoption defined in Equation~\ref{eq:threshold}, determines when the opinion dynamics process effectively begins. As a result, changing this parameter not only directly affects the magnitude of the first wave but also influences the entire pandemic trajectory. Specifically, a higher $\beta$ leads to greater sensitivity to pandemic incidence and, consequently, higher overall adoption rate and a lower peak for each wave. The reverse occurs when $\beta$ is lower.

For the parameter governing the period of past incidence that affects the opinion dynamics system, the baseline model uses a memory horizon with a sigmoid shape that varies over time, whereas the CCF model uses a simple observation window throughout the entire simulation period. The effects of these parameters are complex because they depend on the actual past incidence. In general, a shorter observation window or memory horizon leads to a faster response to recent trends, such as a timely increase in social distancing adoption when incidence surges and a quicker return to low adoption levels after the peak.

The information campaign strength applies only to the CCF model and is intended to reflect the external influence of the information campaign on the system. It directly affects the adoption complexities in both directions, thereby shifting the position of the attractor in the system. The new attractor then pulls the system towards a different overall adoption proportion.

The fatigue rate $\gamma$ applies only to the baseline model and is used in the self-evaluation defined in Equation~\ref{eq:SE}. Although the effect of $\gamma$ theoretically applies throughout the entire simulation, its main influence is on the magnitude of the waves occurring approximately after day 100. A higher $\gamma$ leads to a higher second wave and, where applicable, a higher third wave because a high fatigue rate lowers the self-evaluation value, which in turn results in a lower social distancing adoption rate and a higher number of infections.

A detailed sensitivity analysis is provided in the \nameref{s7}.

\subsection*{National level simulation}

The complex contagion model is initially evaluated against the baseline model using the Australian national dataset, which is the same dataset used in the study that introduced the baseline model~\cite{Chang2025opinion}. Overall, the two models demonstrate comparable performance across the evaluation metrics. However, the CCF model shows a slight improvement in both NRMSE and MDA, as reported in Table~\ref{tab:nrmse_comparison} and Table~\ref{tab:mda_comparison}.

\begin{figure}[htbp]
    \centering
    \includegraphics[width=0.88\linewidth]{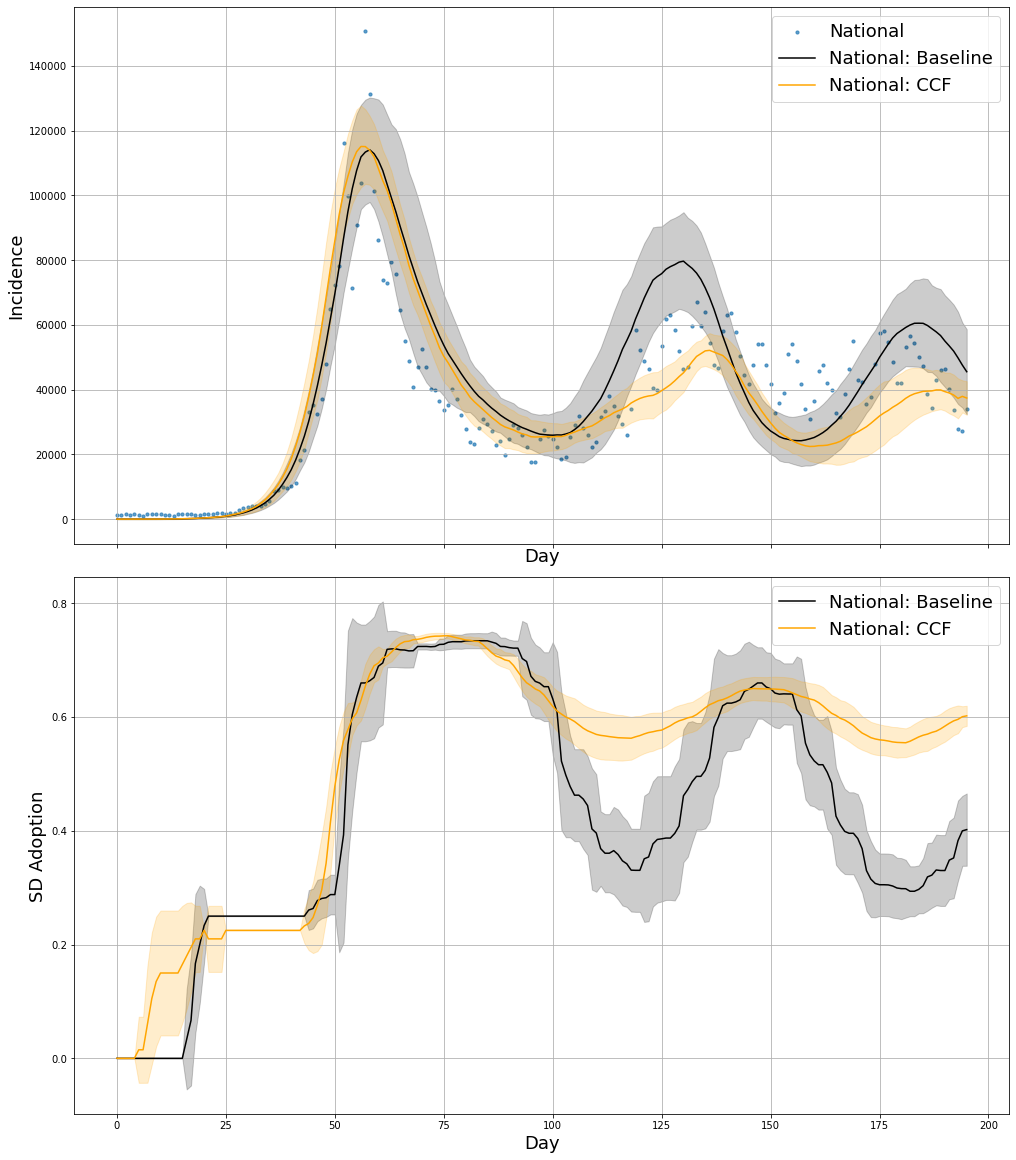}

    \caption{Comparison of simulated incidence and social distance adoption with both baseline and complex contagion model for incidence in Australia national level. Refer to Table \ref{tab:parameter_baseline} and Table \ref{tab:parameter_contagion} for parametrization. Each simulated profile is averaged over 15 runs. Shaded area shows standard deviation. The actual daily incidence between mid-November 2021 and June 2022 is shown in scattered circles.}
    \label{fig:National}
\end{figure}

As illustrated in Figure~\ref{fig:National}, both models show reasonably good performance on the national dataset. In particular, both models accurately reproduce the first wave. For the second and third waves, both models approximately capture the timing and magnitude of the peaks. However, the baseline model tends to slightly overestimate the peak magnitudes, while the CCF model tends to slightly underestimate them.

\subsection*{State level case study: NSW}

For the simulation results for the NSW population, the two models obtain similar evaluation outcomes. The baseline model achieves slightly better performance in terms of NRMSE, as reported in Table~\ref{tab:nrmse_comparison}, while both models obtain very similar level of MDA, as shown in Table~\ref{tab:mda_comparison}.

\begin{figure}[htbp]
    \centering
    \includegraphics[width=0.88\linewidth]{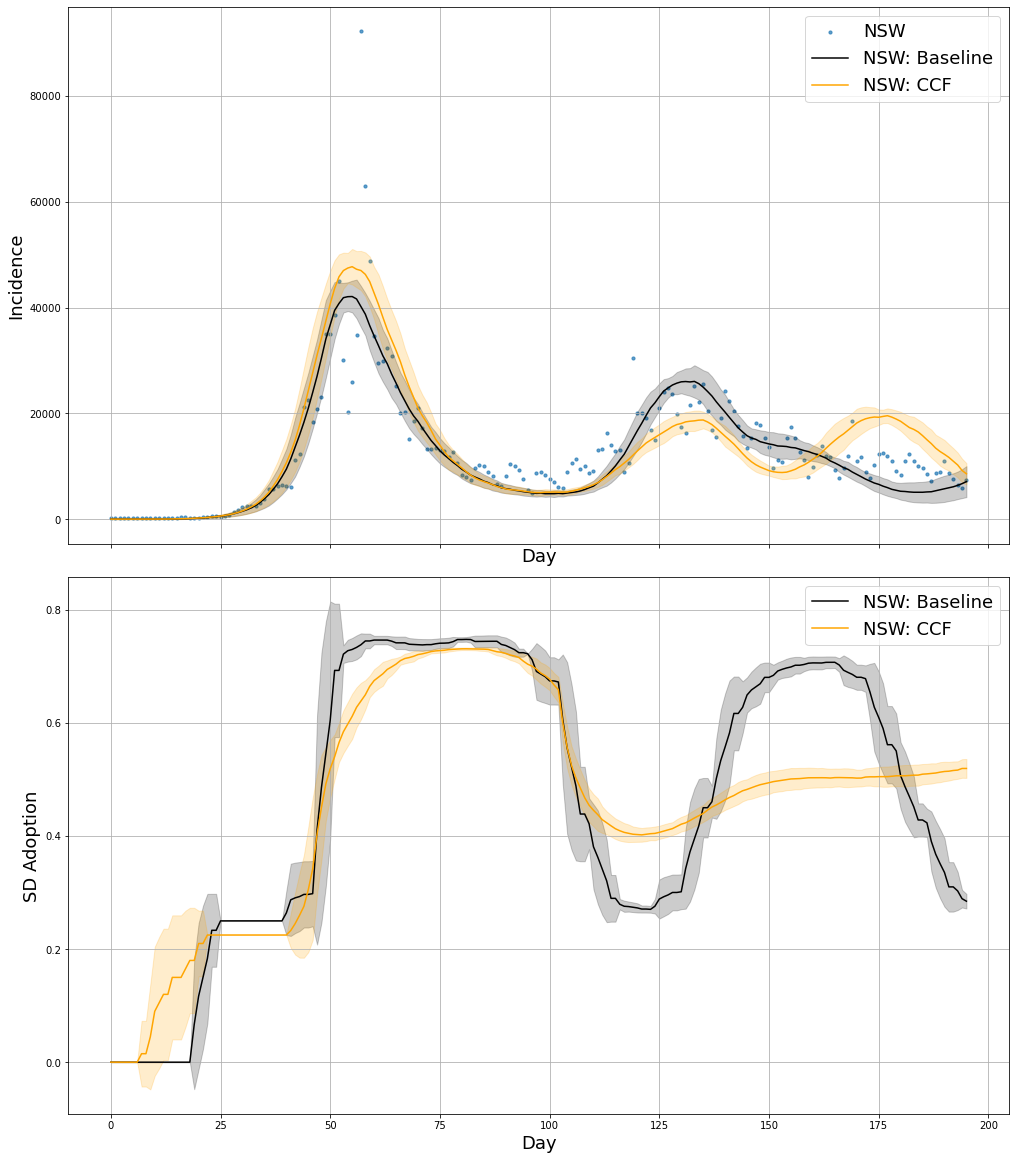}

    \caption{Comparison of simulated incidence and social distance adoption with both baseline and complex contagion model for incidence in NSW. Refer to Table \ref{tab:parameter_baseline} and Table \ref{tab:parameter_contagion} for parametrisation. Each simulated profile is averaged over 15 runs. Shaded area shows standard deviation. The actual daily incidence between mid-November 2021 and June 2022 is shown in scattered circles}
    \label{fig:NSW}
\end{figure}

As shown in Figure~\ref{fig:NSW}, both models reproduce the first and second waves with similar accuracy, although the baseline model performs slightly better in the period immediately after the second peak. Between days 170 and 200, the baseline model slightly underestimates pandemic incidence, while the complex contagion model overestimates it.

\subsection*{State level case study: VIC}

For the simulation results for the VIC population, the two models produced similar evaluation outcomes. The complex contagion model achieved better performance in terms of NRMSE, as reported in Table~\ref{tab:nrmse_comparison}, but performed slightly worse in terms of MDA, as shown in Table~\ref{tab:mda_comparison}.

\begin{figure}[htbp]
    \centering
    \includegraphics[width=0.88\linewidth]{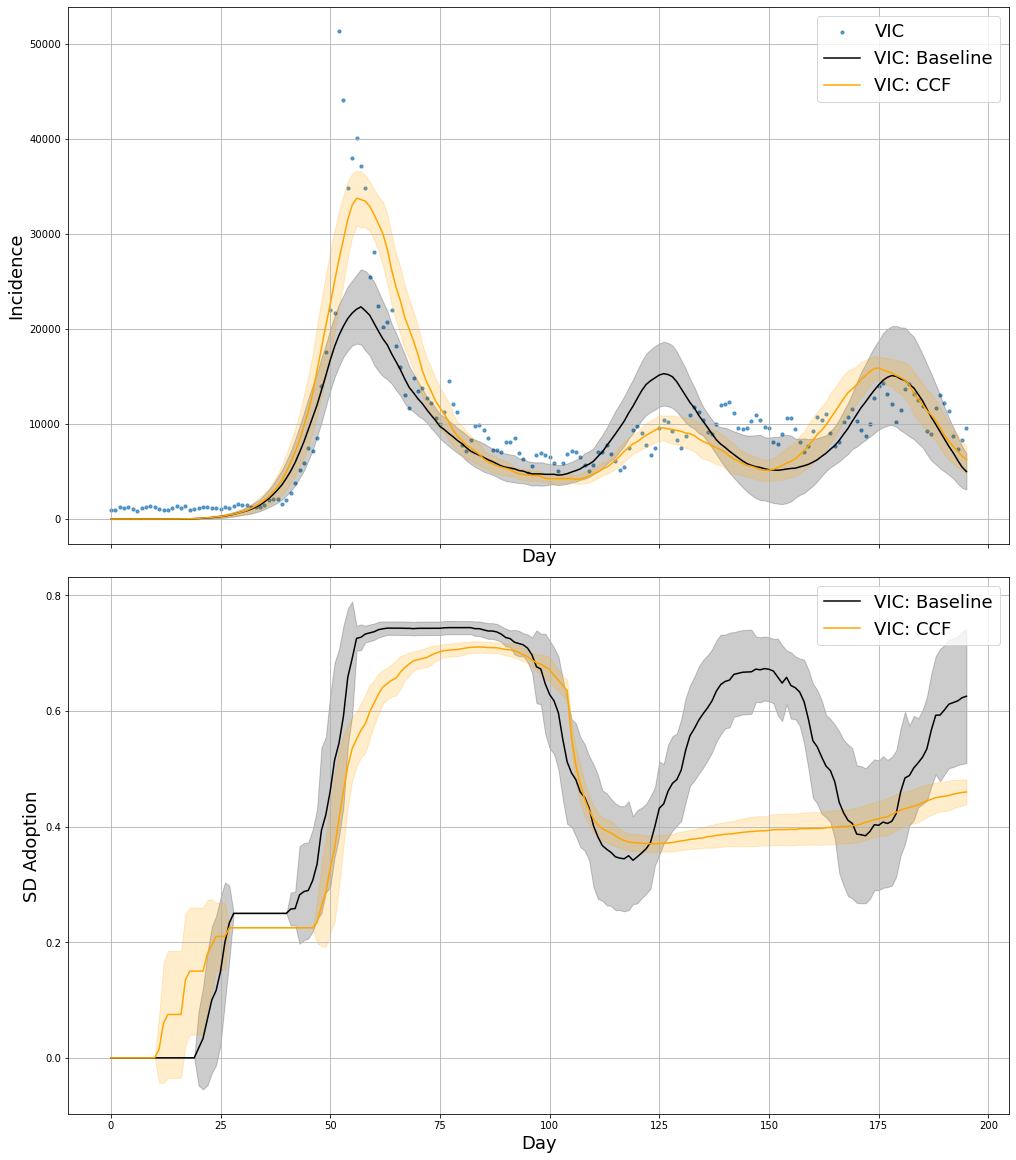}

    \caption{Comparison of simulated incidence and social distance adoption with both baseline and complex contagion model for incidence in VIC. Refer to Table \ref{tab:parameter_baseline} and Table \ref{tab:parameter_contagion} for parametrisation. Each simulated profile is averaged over 15 runs. Shaded area shows standard deviation. The actual daily incidence between mid-November 2021 and June 2022 is shown in scattered circles}
    \label{fig:VIC}
\end{figure}

As shown in Figure~\ref{fig:VIC}, both models captured the overall pattern of pandemic incidence in VIC reasonably well. While the complex contagion model more accurately captured the magnitude of the first peak, the baseline model performed slightly better in reproducing the third wave. In addition, both models slightly missed the timing of the second wave.

\subsection*{State level case study: WA}

The simulation results for Western Australia (WA) reveal an interesting contrast. In the baseline model, the absence of an explicit mechanism for modelling information-campaign-driven exogenous shocks to the opinion dynamics system prevents the model from fully capturing the magnitude of the second wave. This wave is driven by changes in social-distancing-related policy, particularly information campaigns, along with other factors such as waning immunity. For other datasets tested (National, NSW and VIC), the baseline model achieves a very good fit to real-world incidence without explicitly accounting for the impact of information campaigns. This is because the effect of information campaign is likely incorporated implicitly through the shape of the memory horizon and the presence of the fatigue effect. 

According to the evaluation metrics, the CCF model performs substantially better in terms of NRMSE, as shown in Table~\ref{tab:nrmse_comparison}, and slightly better than the baseline model in terms of MDA, as shown in Table~\ref{tab:mda_comparison}.

\begin{figure}[htbp]
    \centering
    \includegraphics[width=0.88\linewidth]{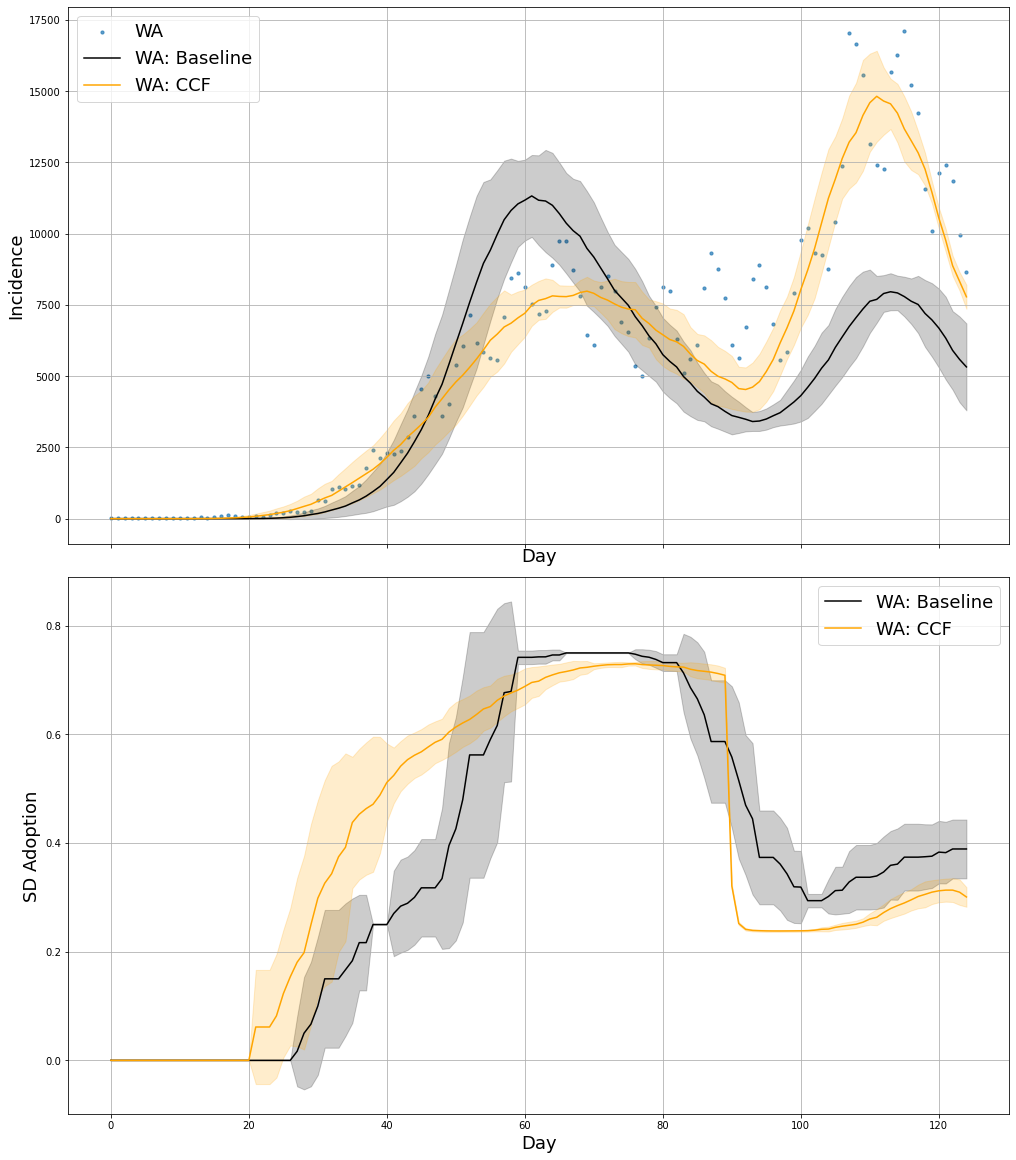}

    \caption{Comparison of simulated incidence and social distance adoption with both baseline and complex contagion model for incidence in WA. Refer to Table \ref{tab:parameter_baseline} and Table \ref{tab:parameter_contagion} for parametrisation. Each simulated profile is averaged over 15 runs. Shaded area shows standard deviation. The actual daily incidence between mid-November 2021 and June 2022 is shown in scattered circles}
    \label{fig:WA}
\end{figure}

Because the baseline model was originally developed and calibrated to reproduce recurrent waves at the national level in Australia~\cite{Chang2025opinion}, it is reasonable that the model adapts well to the two most populous states, NSW and VIC, where social distancing policies are closely synchronised. However, the WA case shows that the baseline model has difficulty when the situation in a specific state differs substantially from the national aggregate, as illustrated in Figure~\ref{fig:WA}. 

For the complex contagion model, a good fit to the real incidence is obtained after the effect of an information campaign is introduced on day 90, reflecting the actual timing of the shift in government policy and public attitudes towards social distancing in the real world. In contrast, the baseline model continues to underestimate the second wave even after extensive parameter tuning.

\section*{Discussion}

\subsection*{Representation of heterogeneous adoption thresholds}

Although the CCF model assigns a common adoption-complexity parameter to each directional transition process, it can represent heterogeneity in the effective adoption thresholds of flexible agents. As shown by Vasconcelos~\emph{et al.}~\cite{10.1098/rsif.2019.0196}, the transition probability defined by Equation~\ref{eq:individual_update}

corresponds to a mean-field description of a population with a distribution of fractional thresholds. Suppose that an agent adopts state $Y$ when the fraction of its neighbours holding $Y$ meets or exceeds a threshold $M\in[0,1]$. The threshold distribution of the population about transition from $X$ to $Y$, $d_{XY}(M)$, can be written in proportional form as $d_{XY}(M)\propto M^{\alpha_{XY}-1}$~\cite{10.1098/rsif.2019.0196}. 

The probability that the threshold does not exceed the observed neighbour fraction is therefore
\begin{equation}
\Pr\left(M\leq\frac{n_i^Y}{z_i}\right)
=
\int_0^{n_i^Y/z_i} d_{XY}(m)\,dm
=
\left(\frac{n_i^Y}{z_i}\right)^{\alpha_{XY}}.
\end{equation}
Thus, $0<\alpha_{XY}<1$ corresponds to a threshold distribution concentrated towards lower values, $\alpha_{XY}=1$ corresponds to a uniform threshold distribution, and $\alpha_{XY}>1$ corresponds to a distribution concentrated towards higher values.

In the CCF model, the time-dependent parameters $\alpha_{AB}(t)$ and $\alpha_{BA}(t)$ can therefore be interpreted as dynamically modifying the effective population-level threshold distributions associated with adopting and discontinuing social distancing. Under this interpretation, crisis feedback and information campaigns alter the shape of the effective threshold distribution rather than imposing one common deterministic threshold on the population. This provides a parsimonious representation of heterogeneous behavioural responsiveness without requiring a separate threshold parameter to be explicitly assigned and calibrated for every flexible agent.

However, its main limitation is that this family does not encompass more general threshold-distribution shapes, such as multimodal distributions. More flexible transition functions could be introduced when empirical evidence supports a richer specification of threshold heterogeneity.

\subsection*{Modelling feedback through dynamic adoption complexity}

In modelling competing behaviours related to social distancing, we replace explicit self-evaluation by individual agents with behaviour-specific adoption complexities at the contagion-process level. Social-distancing adoption and non-adoption are treated as mutually exclusive behaviours that diffuse through social reinforcement. Rather than relying on individual decision rules, the competing complex-contagion framework characterises behavioural dynamics through the structural properties of the diffusion processes, by using different values for adoption complexities~\cite{10.1098/rsif.2019.0196}.

The pandemic incidence trend enters the model as a globally available environmental signal that modifies the relative adoption complexities $\alpha_{AB}$ and $\alpha_{BA}$. These parameters determine the amount of social reinforcement required for behavioural switching. A lower $\alpha_{AB}$ makes behaviour $B$ easier to adopt, whereas a higher value indicates greater resistance to its adoption. Accordingly, a rising pandemic trend lowers $\alpha_{AB}$, facilitating the adoption of social distancing, and raises $\alpha_{BA}$, making non-adoption more difficult to adopt or sustain. This process-level feedback captures the system-wide effect of changing risk conditions without requiring each agent to calculate an individual utility, payoff, or risk assessment.

The baseline and CCF models differ in both the incidence signal used and the persistence of its effects. In the baseline model, each agent calculates the mean incidence over its current memory horizon and uses this value for self-evaluation. In contrast, the CCF model uses the incidence trend over the observation window, as defined in Equation~\ref{eq:incidence_trend}, to update the magnitude and direction of the relative adoption complexities. This trend-based formulation responds to the direction and rate of epidemic change rather than only to its recent level. It can therefore capture the relaxation of protective behaviour when incidence begins to decline, even before absolute incidence becomes low~\cite{Harman2021}.

Overall, this formulation keeps agent-level assumptions minimal by avoiding complicated and time-varying self-evaluation rules, including fatigue rates and memory horizons represented in the baseline model~\cite{Chang2025opinion}, while still preserving agent heterogeneity.

\subsection*{Coupled adoption complexities}

A central design feature of the CCF model is the coupled treatment of the two adoption-complexity parameters, $\alpha_{AB}$ and $\alpha_{BA}$. In the current formulation, the same external feedback signal shifts the two competing contagions in opposite directions: when pandemic incidence is increasing, the complexity of switching from non-adoption to adoption, $\alpha_{AB}$, decreases, while the complexity of switching from adoption to non-adoption, $\alpha_{BA}$, increases. This coupling provides a parsimonious way to represent changes in the relative attractiveness and social legitimacy of two mutually exclusive behaviours. Under worsening crisis conditions, social distancing becomes easier to justify and adopt, while the competing behaviour of non-adoption becomes more difficult to sustain. Conversely, when incidence declines, the perceived need for social distancing weakens and non-adoption becomes easier to adopt. All changes are reflected through the adoption complexity parameter, instead of using individual-level self-evaluation. This coupling is therefore aligned with the intended role of the model: to capture the system-level competition between behavioural alternatives without introducing separate individual-level decision rules for each transition.

The current CCF framework assumes strict coupling between adoption and discontinuation. While in reality these processes may respond differently to the same external signals. A potential future extension is to assign separate sensitivity parameters to $\alpha_{AB}$ and $\alpha_{BA}$, allowing asymmetric behavioural response and inertia to be represented.

\section*{Conclusion}
We introduced a competing complex contagion with feedback (CCF) model that couples socially reinforced behavioural change with the evolving state of a crisis. In the pandemic application, recent incidence trends modify the relative complexity of adopting and abandoning social distancing, while changes in social-distancing adoption subsequently influence disease transmission. The model also represents information campaigns as explicit interventions that shift the relative adoption complexity of the competing behaviours. This formulation provides a parsimonious alternative to modelling individual risk evaluation, memory horizon, and behavioural fatigue as separate mechanisms.

The mean-field analysis of the CCF model demonstrates that cumulative feedback determines how the equilibrium shifts between the two opinions, while inflexible agents prevent the system from reaching complete consensus and produce an interior equilibrium. The agreement between the frozen-equilibrium curve and the simulation results indicates that the mean-field formulation provides a useful approximation of the CCF dynamics.

As a validation case study, the CCF model was integrated with a census-calibrated agent-based model of COVID-19 transmission and reproduced recurrent epidemic waves across the Australian national, NSW, VIC, and WA datasets. Its performance was broadly comparable to that of the baseline opinion-dynamics model for the national, NSW, and VIC datasets, while providing a substantially better fit for WA, where an explicit information-campaign mechanism was important for representing a marked change in behavioural and policy conditions. These results indicate that recurrent waves can emerge from the feedback between epidemic trends, social reinforcement, and behavioural adoption without requiring a highly parameterised individual-level decision process.

Overall, the CCF framework offers an interpretable and transferable approach to modelling co-evolving crisis and behavioural dynamics. Although further validation is required across different populations, network structures, crisis settings, and behavioural data, the model provides a useful foundation for examining how social reinforcement, environmental feedback, and external interventions jointly shape collective responses~\cite{Thompson2022}. Beyond pandemic modelling, the framework may be applicable to other settings in which competing behaviours or opinions both respond to and modify an evolving external system.

\bibliographystyle{ieeetr}
\bibliography{ref}

\section*{Supporting information}\label{Appendix}
\paragraph*{S1 Appendix. Agent-based model augmented with social network}\label{s1}

We adopt the census-calibrated agent-based model (ABM) developed by Chang~et~al.~\cite{amtrac_2020}, which was subsequently extended to model the Omicron variant of SARS-CoV-2~\cite{Chang2023persistence} and later used as the pandemic model in an opinion-coupled simulation~\cite{Chang2025opinion}. The model consists of approximately 25.4~million anonymous agents, each assigned demographic attributes, including age, gender, residential area, and workforce or educational group, based on data from the 2021 Australian Census~\cite{Census_2021}. These attributes define the social contexts in which disease transmission occurs~\cite{Chang2025opinion}.

Transmission occurs in discrete time steps, with each agent transitioning among the states \emph{susceptible}, \emph{latent}, \emph{infectious} (asymptomatic or symptomatic), and \emph{recovered}. Infections are seeded in areas near international airports and spread through two interaction phases on each weekday: a \emph{daytime} cycle covering workplace and educational contacts, and a \emph{night-time} cycle limited to residential contexts, including households, household clusters, neighbourhoods, and communities. Weekends consist of two night-time cycles~\cite{amtrac_2020}.

The probability that an exposed agent becomes infected is affected by prior vaccination and the adoption of non-pharmaceutical interventions (NPIs). Only a fraction of infections are detected, with the detection rate calibrated using the prevalence of anti-nucleocapsid antibodies in the Australian population, consistent with the voluntary self-reporting system during the Omicron stage~\cite{Chang2025opinion}. Following previous studies~\cite{Chang2023persistence, Zachreson2021, Nguyen2022}, we initialise the population with $90\%$ pre-emptive vaccination coverage across two vaccine types (priority and general, of higher and lower efficacy, respectively) and track immune waning at the individual level, allowing recovered agents to be reinfected~\cite{Chang2025opinion}.

Four NPIs are modelled: \emph{case isolation}~(symptomatic and detected asymptomatic infectious agents), \emph{home quarantine}~(household members of confirmed cases), \emph{social distancing}~(SD, susceptible agents), and \emph{school closures}~(school-aged agents, their households, and teachers). Each NPI reduces interaction strength in the relevant contexts. All NPIs remain static throughout the simulation except SD, which we model as a voluntary, time-varying decision that captures the joint effects of reduced mobility, mask-wearing, and physical distancing. Its adoption is driven by complex contagion dynamics.

To construct the social network, we adopt the simple lattice configuration model, inspired by the configuration model~\cite{Newman2001} and the Watts-Strogatz model~\cite{Watts_Strogatz_1998}. This hybrid approach has been validated in our previous study~\cite{Huang2025}.

\paragraph*{S2 Appendix. Other implementation details}\label{s2}
To initialise the complex contagion model, we introduce a daily incidence starting threshold representing the point in time at which the broader population begins to perceive the pandemic as a salient personal risk. This assumption is motivated by a retrospective study of social-distancing adoption in Australia during the Omicron phase, which found that adoption remained relatively low for an extended period before the first peak and shifted to a higher level only after national incidence had reached substantially elevated levels~\cite{Chang2023persistence}.

Accordingly, we assume that the complex contagion process is not triggered among the general population before daily incidence crosses the specified threshold. During this initial phase, only a fixed proportion of inflexible agents adhere to social distancing. These agents maintain a pro-adoption opinion independently of prevailing social influence and current incidence levels.

Once daily incidence exceeds the threshold, all remaining agents become aware of the pandemic and the opinion-dynamics process is activated. At this point, the inflexible agents serve as the initial seeds of pro-adoption opinion, allowing social-distancing behaviour to spread through complex contagion within the broader population.

Besides, the update process occurs every $\Delta t$ time units, with $\Delta t$ set to 7 in this study, which is the same value as the social distancing opinion update interval in the baseline model~\cite{Chang2025opinion}.

\paragraph*{S3 Appendix. Details for simulation parametrisation}\label{s3}
In the state-level simulations, most model parameters are held constant. To assess the model’s performance across different scenarios, adjustments are limited to variables considered likely to differ meaningfully between states. Parameters in both models are also selected to represent broadly analogous underlying constructs, with the aim of improving the comparability of the two models.

For the baseline model, four variables are adjusted: the fatigue rate $\gamma$, the perceived infection rate $\beta$, and the two parameters governing the shape of the memory horizon. This choice reflects expected differences among states in both the severity of their pandemic situations and the stringency with which pandemic-related policies were enforced. The memory horizon is interpreted as indirectly representing the potential effects of policy changes. Specifically, the day offset (delay) $L$ is used as a proxy for the timing of policy implementation, while the height of the memory horizon $u$ is intended to represent variation in enforcement strength. The perceived infection rate $\beta$ is intended to account, at least partly, for differences in states’ prior pandemic histories. Thus, even when two states have the same incidence proportion relative to their total populations, residents may respond differently depending on their past experiences. For example, a state that experienced several significant outbreaks before the Omicron period may exhibit lower sensitivity to a given incidence level. The fatigue rate $\gamma$ serves a similar purpose and is intended to represent potential variation in pandemic fatigue across states. It is therefore also varied in each state-level fitting.

For the complex contagion model, functionally analogous parameters are selected wherever possible to improve consistency and comparability with the baseline model. The window size is treated as analogous to the memory horizon in the baseline model. The policy strength parameter provides a simplified representation of the influence of state-level policy on behaviour. Although policy is not modelled explicitly in the opinion dynamics module of the baseline model, we interpret its influence as being represented indirectly through the shape of the memory horizon. A third parameter, the starting threshold of the opinion dynamics module, specifies the point at which individuals are assumed to begin perceiving infection risk and engaging in opinion dynamics related to SD adoption. This parameter is treated as functionally analogous to the perceived infection rate in the baseline model. Together with the internal opinion threshold for adopting SD, the perceived infection rate serves a broadly similar functional role to that of the starting threshold in the complex contagion model.

It is worth noting that four free parameters are selected for the baseline model, whereas only three are selected for the complex contagion model. We consider this asymmetry reasonable for two primary reasons. First, despite the structural differences between the two models, we aim to select free parameters that are as functionally comparable as possible, thereby supporting a more meaningful comparison. Second, the choice of three parameters for the CCF model is primarily due to the model’s structural constraints rather than by an arbitrary decision. The model contains relatively few parameters that can be meaningfully adjusted in the transfer setting, and these three represent the main parameters available for adjustment.

\paragraph*{S4 Appendix. Evaluation metrics: normalised root mean square error}\label{s4}

Model accuracy is primarily evaluated using the Normalized Root Mean Square Error (NRMSE), which measures prediction error relative to the observed mean, making it dimensionless and comparable across variables with different scales and units.

Its squared-error formulation inherently penalizes large deviations more heavily than small ones, making it well suited to time-series evaluation when substantial mispredictions, such as those in pandemic incidence, may have greater practical consequences than suggested by average percentage error alone.

NRMSE is computed as the root mean square error normalized by the mean of the observations, providing a scale-independent measure of the deviation between simulated and observed trajectories. It serves as the primary evaluation metric because it is sensitive to both phase misalignment and magnitude discrepancy and provides a single, interpretable summary of overall trajectory agreement. Furthermore, the open-loop nature of the simulation framework makes NRMSE a stringent benchmark because the model does not correct accumulated drift during simulation.

\paragraph*{S5 Appendix. Evaluation metrics: modified mean directional accuracy}\label{s5}

To complement NRMSE, which is insensitive to the directional structure of the epidemic trajectory, a modified Mean Directional Accuracy (MDA) is used, adapted from its standard form in three respects.

Formally, for a smoothed observed series $\tilde{y}$ and predicted series $\hat{y}$, the modified MDA over horizon $h$ is defined as:
\begin{equation}
    \text{MDA}^* = \frac{100}{|\mathcal{M}|} \sum_{t \in \mathcal{M}}
    \mathbf{1}\!\left[
        \operatorname{sgn}\!\left(\tilde{y}_{t+h} - \tilde{y}_{t}\right)
        = \operatorname{sgn}\!\left(\hat{y}_{t+h} - \hat{y}_{t}\right)
    \right]
    \label{eq:mda_modified}
\end{equation}
\noindent where $\mathcal{M} = \{t : |\tilde{y}_{t+h} - \tilde{y}_{t}| > \tau\}$ is the set of time points at which the observed trend exceeds a noise threshold.

Three adaptations were made to handle noisy real world pandemic data. First, $\tilde{y}$ is obtained by smoothing the observed series with a bidirectional EWMA (span = 7, matching the weekly reporting cycle) prior to computing deltas, since a causal EWMA would introduce phase lag that displaces turning points and deflates accuracy near wave onset or peak. Second, evaluation is restricted to $\mathcal{M}$, the set of trend-active points exceeding the noise floor, so that flat or noise-dominated periods do not contribute spurious directional agreement. Third, deltas are computed over a multi-step horizon ($h = 7$ days) rather than single-step differences, since day-to-day changes are dominated by reporting variance and carry little signal about the underlying trend.

\paragraph{S6 Appendix. Sensitivity analysis}\label{s7}
For the sensitivity analysis, we vary one parameter at a time while holding all other parameters fixed. The detailed parameter configurations are summarized in Table \ref{tab:configuration_parameters_SA}. We use NSW as a representative example, and the effects of the parameters are similar across the different datasets.

\begin{table}[ht]
\centering
\caption{Configuration parameters for sensitivity analysis. $h$ denotes the length of the observation window, and $\theta$ denotes the strength of the information campaign}
\label{tab:configuration_parameters_SA}
\begin{tabular}{|c|c|c|c|}
\hline
\textbf{Configuration Name} & \textbf{Starting threshold} & \textbf{$h$} & \textbf{$\theta$} \\ \hline
W1 & 20000 & 28 & -1500 \\ \hline
W2 & 20000 & 35 & -1500 \\ \hline
IC1 & 20000 & 42 & -1000 \\ \hline
IC3 & 20000 & 42 & -2000 \\ \hline
W3/IC2 & 20000 & 42 & -1500 \\ \hline

\end{tabular}
\end{table}

\begin{figure}[htbp]
\centering
\includegraphics[width=0.88\linewidth]{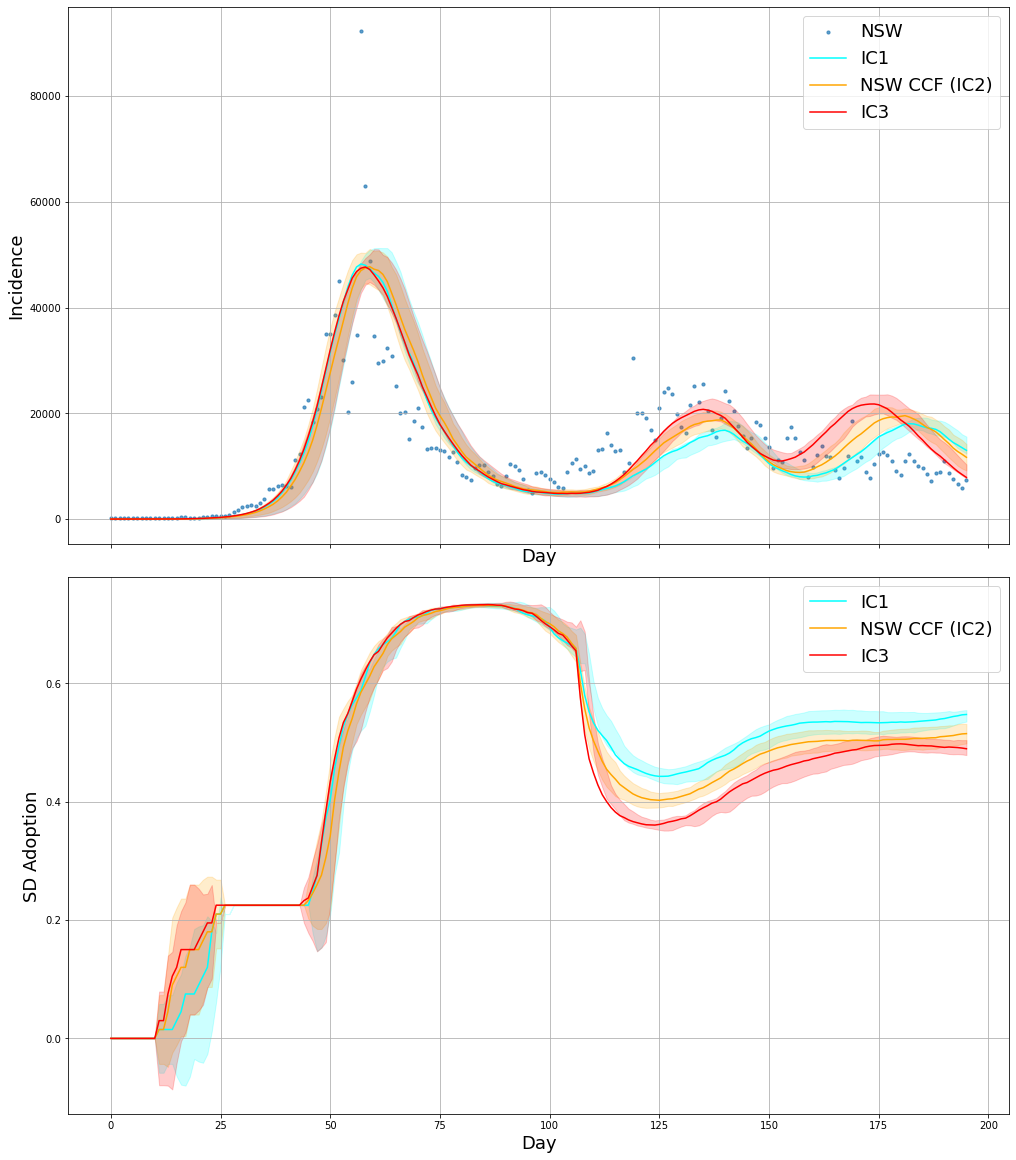}
\caption{Comparison of the incidence and social distance adoption simulated by the complex contagion model under different values of the strength of the information campaign for NSW. The parameter configurations are provided in Table \ref{tab:configuration_parameters_SA}. Each simulated profile is averaged over 15 runs. The observed daily incidence from mid-November 2021 to June 2022 is shown as circles.}
\label{fig:IC}
\end{figure}

We first examine the effect of changing the strength of the information campaign. The information campaign is triggered only on the specified date corresponding to the real-world event, which is day 105 for NSW. The strength of the information campaign determines the magnitude of the change in the SD adoption rate in the direction promoted by the campaign. As shown in Figure \ref{fig:IC}, a greater strength therefore produces a larger change in the SD adoption rate after information campaign is triggered.

\begin{figure}[htbp]
\centering
\includegraphics[width=0.8\linewidth]{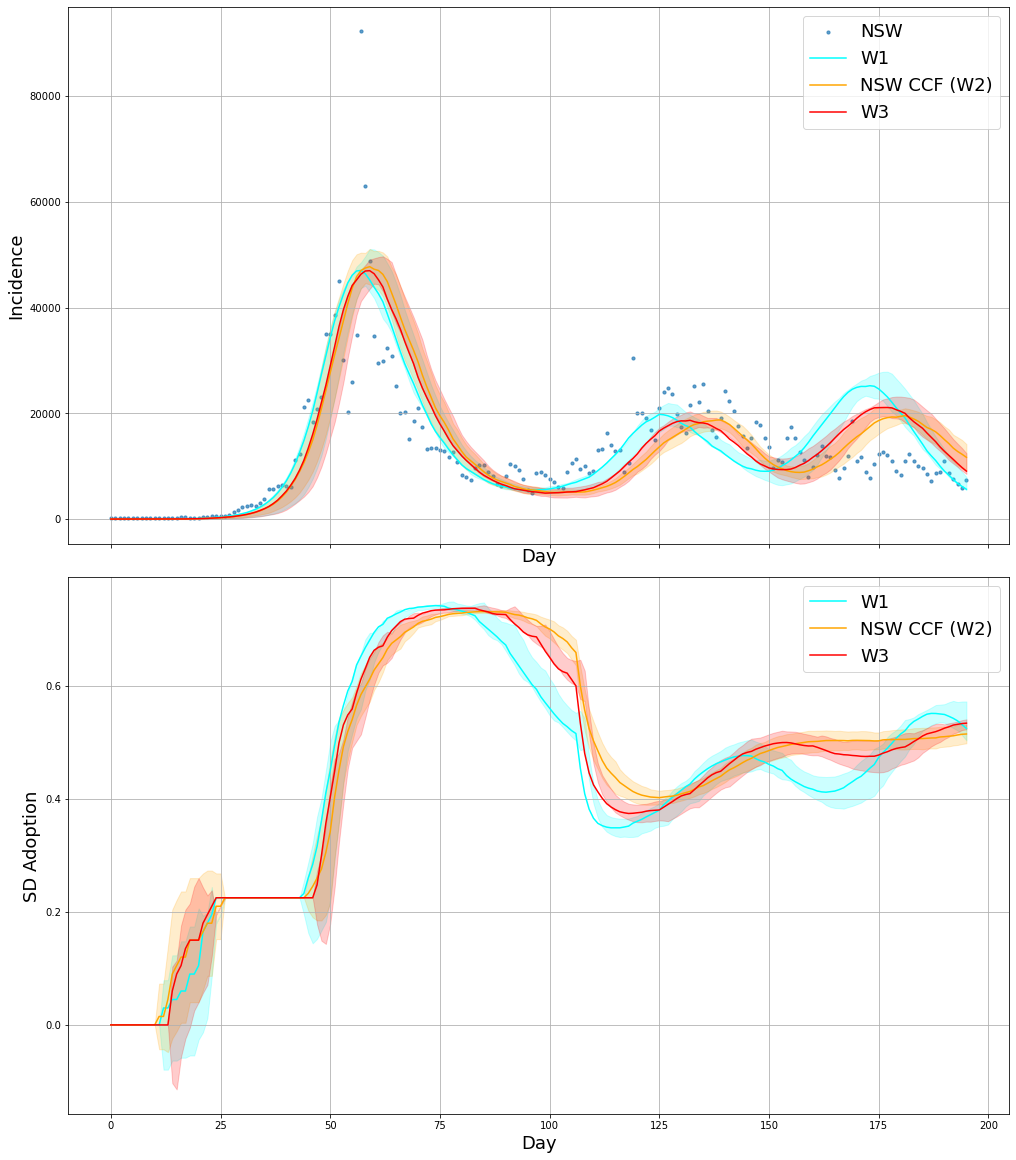}

\caption{Comparison of the incidence and social distance adoption simulated by the complex contagion model under different values of the length of the observation window for NSW. The parameter configurations are provided in Table \ref{tab:configuration_parameters_SA}. Each simulated profile is averaged over 15 runs. The observed daily incidence from mid-November 2021 to June 2022 is shown as circles.}
\label{fig:Window}
\end{figure}

The effect of changing the length of the observation window is more complex. In general, a shorter length of the observation window results in more volatile SD adoption and greater sensitivity to recent trends. As shown in Figure \ref{fig:Window}, configuration W1 exhibits more fluctuations than the other configurations. By contrast, configurations with a longer length of the observation window produce more stable patterns of SD adoption over time.

\end{document}